\documentclass[conference,10pt]{IEEEtran}

\usepackage[utf8]{inputenc}
\usepackage{graphics,graphicx,color,epsfig}
\usepackage{amsmath,amssymb,amsthm,amsfonts,bm,bbm}
\usepackage{algorithmicx,algorithm,algpseudocode}
\usepackage{multirow}
\usepackage{cite}
\usepackage[normalem]{ulem}

\usepackage{tikz}
\usetikzlibrary{automata, positioning, arrows,shapes.multipart,fit}
\usepackage{pgfplots}

\usepackage{subcaption}
\usepackage{todonotes}
\presetkeys{todonotes}{color=blue!20}{}

\title{Blind decoding in $\alpha$-Stable noise: An online learning approach}
\author{Vishnu Raj \hspace{20pt} Sheetal Kalyani\\
   \hspace{-0 cm}Department of Electrical Engineering,\\ Indian Institute of Technology Madras, \\
   Chennai, India, 600 036. \\
   \texttt{\{ee14d213,skalyani\}@ee.iitm.ac.in}
}

\begin{document}
    \maketitle
    
    \begin{abstract}
        A novel method for performing error control coding in Symmetric
        $\alpha-$Stable noise environments without any prior knowledge about the value of $\alpha$
        is introduced.
        We use an online learning framework which employs multiple distributions to decode 
        the received block and then combines these results based on the past performance of each 
        individual distributions. 
        The proposed method is also able to
        handle a mixture of Symmetric $\alpha-$Stable distributed noises.
        Performance results in turbo coded system highlight the utility of the work.
        
    \end{abstract}

    
    \section{Introduction}
Impulsive noise which is commonly found in modern communication systems can be modeled by
Symmetric $\alpha-$Stable ($S \alpha S$) distributions \cite{tsakalides1995array}. 
$S \alpha S$ distribution is found to well model the aggregate interference found in cognitive 
radio networks, 
turbo-coded OFDM systems \cite{kalyani2008interference}, 
ultra-wideband ad-hoc networks \cite{el2010alpha}, 
multi-user interference \cite{pinto2010communication}, etc.
This ubiquitous nature of impulsive noise in wireless communication systems and the capability
of $S\alpha S$ distribution to capture the heavy-tailed behavior of observed impulsiveness have
renewed the research interest in stable distributions. 

The knowledge of noise distribution along with the exact parameters is crucial in iterative
error correction coding; mis-specification of the noise distribution can severely degrade the
performance. 
Except for the special cases of Gaussian
distribution ($\alpha = 2$) and Cauchy distribution ($\alpha = 1$), simple closed-form expression for 
the general case of $S \alpha S$ distribution is not available.
This led to approximation based methods for calculating the density functions
\cite{gu2012decoding,dimanche2014detection}, 
where the impulsiveness parameter
$\alpha$ is assumed to be known apriori.
In \cite{mestrah2018blind}, a supervised learning based method for estimating the parameters
for the approximation proposed in \cite{dimanche2014detection} is developed. 
Estimating the impulsive behavior of heavy-tailed
noise or the corresponding $\alpha$ parameter of the distribution requires a substantial number of
observations 
\cite{mohammadi2015estimating}. 
It could also result in discarding blocks
of data without decoding to estimate the channel impulsiveness or failure to decode the transmitted
blocks due to incorrect noise parameters. Hence, in the presence of $S\alpha S$ noise where $\alpha$
is unknown, it remains a challenge to apply iterative error correction coding.

In this paper, we look at the reception of turbo coded blocks in the presence of unknown channel
noise. First, we evaluate multiple possible noise models simultaneously using Maximum a Posteriori 
probability (MAP) decoders. Then we employ Hedge\cite{freund1997decision}, a popular
online learning method, to combine the information from multiple decoders in a game theoretic 
fashion to incur least decoding error. The main contributions of this work are
\begin{enumerate}
    \item Proposes a multi-pair MAP decoder architecture and an online learning based technique
            for performing turbo decoding without any prior knowledge about channel parameter $\alpha$.
    \item Shows that the proposed scheme is able to match the performance of the optimal decoder, 
            which has full knowledge of $\alpha$, in the case of both single and mixture of multiple
            noise distributions.
    \item Gives a low complexity version of the proposed approach.
\end{enumerate}
To summarize, we present a practical solution for 
performing iterative error correction coding when the family of possible noise distributions is
known, but the exact parameters of the distribution are unknown.
    \section{Proposed Approach}
Consider a linear discrete time memory-less transmission channel with additive impulsive 
noise modelled using 
$S \alpha S$ distribution. $S \alpha S$
distribution is defined by the characteristic function \cite{tsakalides1995array},
\begin{align}
    \phi(\omega) = \exp( - \gamma |\omega|^\alpha ), \qquad -\infty < \omega < \infty,   \label{eqn:sas_cf}
\end{align}
where $0 < \alpha \leq 2$ is the characteristic exponent related to the impulsiveness of the
random variable
(lower the $\alpha$, higher is the impulsiveness)
and $\gamma > 0$ is the dispersion. 
Two special cases of $S \alpha S$ distributions are Cauchy distribution (with $\alpha = 1$) defined 
as $Cauchy(\gamma) := S\alpha S(1,\gamma)$ and Gaussian distribution (with $\alpha = 2$) defined as
$\mathcal{N}(0,\sigma^2) := S\alpha S(2,\sigma^2/2)$.

For $\alpha < 2$, the variance of $S \alpha S$ random variables is not defined. Hence,
instead of the traditional Signal-to-Noise power ratio (SNR) measure, 
an alternative measure termed as Geometric 
SNR (GSNR) was proposed in \cite{chuah2000nonlinear} and is defined as
\begin{align}
    GSNR = \frac{1}{2 C_g} \left( \frac{A_k}{S_0} \right)^2.
\end{align}
$S_0 = \frac{(C_g \gamma)^{1/\alpha}}{C_g}$ is the geometric power of heavy tailed
noise and  $C_g \approx 1.78$ is the exponential of the Euler constant. 

We consider a
turbo encoder that takes in a block of $K$ data bits and outputs a
block of $N > K$ bits. The coded block of data is then transmitted through the channel and
gets corrupted by unknown noise signals. Let
$\textbf{x}$ denote the transmitted codeword (of block length $N$) and let $\textbf{y}$ denote
the received codeword. Each bit in the $N$-length block is independently corrupted by impulsive
noise, $z$, as 
\begin{align}
    y[n] &= x[n] + z[n], 
            \qquad n = 1,\ldots,N,
\end{align}
where $z[n] \sim S\alpha S(\alpha,\gamma)$. The turbo decoder 
takes in $N$
length received vector $\textbf{y}$ and decodes the data bits of length $K$. Traditional
turbo decoders employ a pair of Bahl-Cocke-Jelinek-Raviv (BCJR) decoders. Decoding
proceeds in iterations in which each BCJR decoder calculates the Log-Likelihood Ratio (LLR) as
\begin{align}
    L(x_k|\textbf{y}) = \ln 
                            \frac{\mathbb{P}(x_k=1|\textbf{y})}
                            {\mathbb{P}(x_k=0|\textbf{y})},  \label{eqn:llr}
\end{align}
for each of the databit $x_k, k = 1,\ldots,K$ and passing the \textit{extrinsic} information 
to its partner decoder. 

\subsection{Approximating $S\alpha S$ density}
For $S \alpha S$ channel, 
computing LLR is a challenge due to the unavailability of a closed-form expression for evaluating
density except for the case of Cauchy and Gaussian noises. When $\alpha$ is known a priori, 
the work in \cite{li2008approximate}  proposed to use a mixture of Gaussian and
Cauchy densities to approximate the $S\alpha S$ density (for $1 \leq \alpha \leq 2$) as
\begin{align}
    f_{\alpha}(x) = \epsilon_\alpha \frac{\gamma}{\pi (x^2+\gamma^2)} + 
                    (1-\epsilon_\alpha) \frac{1}{2 \gamma\sqrt{\pi}} \exp\left( -\frac{x^2}{4\gamma^2} \right),    \label{eqn:bcgm_model}
\end{align}
where $\epsilon_\alpha = \frac{4-\alpha^2}{3\alpha^2}$. This Bi-parameter 
Cauchy Gaussian Mixture (BCGM) model
is analyzed in a minimum error setting in \cite{xu2012minimum} (abbreviated as MEBCGM) and is shown that
the minimum error in approximation can be obtained by setting $\epsilon_\alpha = \frac{B_\alpha}{A}$
where $B_\alpha = \int \limits_{0}^{\infty} \left( e^{-\omega-\omega^\alpha} + e^{-2\omega^2} -
e^{-\omega-\omega^2} - e^{-\omega^\alpha-\omega^2}\right) d\omega $ and $A = \int \limits_{0}^{\infty}
\left( e^{-2\omega} + e^{-2\omega^2} - 2e^{-\omega-\omega^2} \right) d\omega$. 
This expression is further simplified to get a quadratic fit for minimum error as
\begin{align}
    \epsilon_{\alpha} = 3.01753 - 2.53103\alpha + 0.513504\alpha^2. \label{eqn:eps_quad_fit}
\end{align}
We propose to use 
(\ref{eqn:bcgm_model}) with (\ref{eqn:eps_quad_fit}), for MAP decoder pairs with 
different $\alpha$ values for obtaining branch
transition probabilities and the corresponding approximate LLRs.

\subsection{Multi-pair MAP decoders and Online Combining}

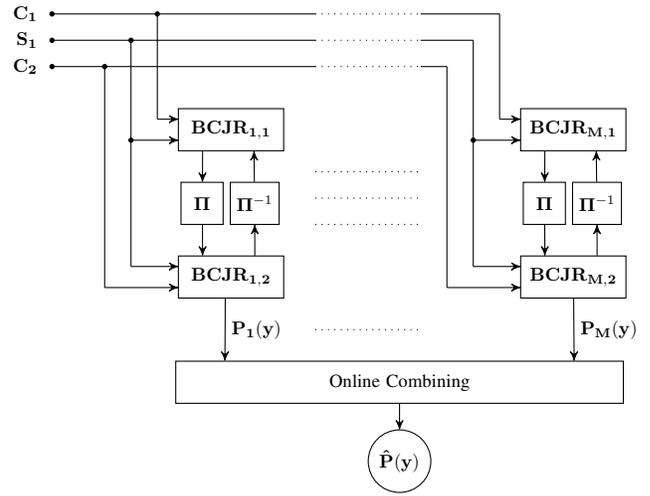
\begin{figure}[t]
    \centering
    \begin{tikzpicture}[auto,>=stealth',every text node part/.style={align=center},scale=0.70, every node/.style={transform shape}]
    \begin{scope}[auto,node distance=1.0cm]
        \draw (0,-0.0) coordinate (c1);    \filldraw (c1) circle(0.4mm) node[left=1.5mm,fill=white]{$\mathbf{C_1}$};
        \draw (0,-0.5) coordinate (s1);    \filldraw (s1) circle(0.4mm) node[left=1.5mm,fill=white]{$\mathbf{S_1}$};
        \draw (0,-1.0) coordinate (c2);    \filldraw (c2) circle(0.4mm) node[left=1.5mm,fill=white]{$\mathbf{C_2}$};
        
        
        \draw (2.0,-0.0) coordinate (c1_11);    \filldraw (c1_11) circle(0.4mm);    \draw (c1) -- (c1_11);
        \draw (1.5,-0.5) coordinate (s1_11);    \filldraw (s1_11) circle(0.4mm);    \draw (s1) -- (s1_11);
        \draw (1.0,-1.0) coordinate (c2_11);    \filldraw (c2_11) circle(0.4mm);    \draw (c2) -- (c2_11);
        
        \draw (1.5,-2.4) coordinate (s1_12);    \draw (s1_11) -- (s1_12);
        \draw (2.0,-2.0) coordinate (c1_12);    \filldraw (s1_12) circle(0.4mm);    \draw (c1_11) -- (c1_12);
        
        \draw (1.5,-4.8) coordinate (s1_13);    \draw (s1_12) -- (s1_13);
        \draw (1.0,-5.2) coordinate (c2_13);    \draw (c2_11) -- (c2_13);
        
        \node [draw,fill=white,rectangle,minimum height=0.8cm,minimum width=2.0cm] at (3.4,-2.2) (bcjr11) {$\mathbf{BCJR_{1,1}}$};
        \draw [->] (s1_12) -- (2.4,-2.4);
        \draw [->] (c1_12) -- (2.4,-2.0);
        \node [draw,fill=white,rectangle,minimum height=0.8cm,minimum width=0.8cm,below of=bcjr11,xshift=-0.55cm,yshift=-0.4cm] (pi11) {$\mathbf{\Pi}$};
        \draw [->] (bcjr11.217) -- (pi11.89);
        \node [draw,fill=white,rectangle,minimum height=0.8cm,minimum width=0.8cm,right of=pi11] (pi12) {$\mathbf{\Pi}^{-1}$};
        \draw [->] (pi12.91) -- (bcjr11.317);
        \node [draw,fill=white,rectangle,minimum height=0.8cm,minimum width=2.0cm,below of=bcjr11,yshift=-1.8cm] (bcjr12) {$\mathbf{BCJR_{1,2}}$};
        \draw [->] (s1_13) -- (2.4,-4.8);
        \draw [->] (c2_13) -- (2.4,-5.2);
        \draw [->] (pi11.271) -- (bcjr12.143);
        \draw [->] (bcjr12.42) -- (pi12.270);
        
        \draw (c1_11) -- (5.0,-0.0);
        \draw (s1_11) -- (5.0,-0.5);
        \draw (c2_11) -- (5.0,-1.0);
        
        \draw[dotted] (5.0,-0.0) -- (7.0,-0.0);
        \draw[dotted] (5.0,-0.5) -- (7.0,-0.5);
        \draw[dotted] (5.0,-1.0) -- (7.0,-1.0);
        
        \draw[dotted] (5.0,-3.0) -- (7.0,-3.0);
        \draw[dotted] (5.0,-3.5) -- (7.0,-3.5);
        \draw[dotted] (5.0,-4.0) -- (7.0,-4.0);
        
        \draw[dotted] (5.0,-6.0) -- (7.0,-6.0);
        
        \draw (8.5,-0.0) coordinate (c1_k1);    \draw (7.0,-0.0) -- (c1_k1);
        \draw (8.0,-0.5) coordinate (s1_k1);    \draw (7.0,-0.5) -- (s1_k1);
        \draw (7.5,-1.0) coordinate (c2_k1);    \draw (7.0,-1.0) -- (c2_k1);
        
        \draw (8.0,-2.4) coordinate (s1_k2);    \draw (s1_k1) -- (s1_k2);
        \draw (8.5,-2.0) coordinate (c1_k2);    \filldraw (s1_k2) circle(0.4mm);    \draw (c1_k1) -- (c1_k2);
        
        \draw (8.0,-4.8) coordinate (s1_k3);    \draw (s1_k2) -- (s1_k3);
        \draw (7.5,-5.2) coordinate (c2_k3);    \draw (c2_k1) -- (c2_k3);
        
        \node [draw,fill=white,rectangle,minimum height=0.8cm,minimum width=2.0cm] at (9.9,-2.2) (bcjrk1) {$\mathbf{BCJR_{M,1}}$};
        \draw [->] (s1_k2) -- (8.9,-2.4);
        \draw [->] (c1_k2) -- (8.9,-2.0);
        
        \node [draw,fill=white,rectangle,minimum height=0.8cm,minimum width=0.8cm,below of=bcjrk1,xshift=-0.55cm,yshift=-0.4cm] (pik1) {$\mathbf{\Pi}$};
        \draw [->] (bcjrk1.217) -- (pik1.89);
        \node [draw,fill=white,rectangle,minimum height=0.8cm,minimum width=0.8cm,right of=pik1] (pik2) {$\mathbf{\Pi}^{-1}$};
        \draw [->] (pik2.91) -- (bcjrk1.317);
        \node [draw,fill=white,rectangle,minimum height=0.8cm,minimum width=2.0cm,below of=bcjrk1,yshift=-1.8cm] (bcjrk2) {$\mathbf{BCJR_{M,2}}$};
        \draw [->] (s1_k3) -- (8.9,-4.8);
        \draw [->] (c2_k3) -- (8.9,-5.2);
        \draw [->] (pik1.271) -- (bcjrk2.143);
        \draw [->] (bcjrk2.42) -- (pik2.270);
        
        \node [draw,fill=white,rectangle,minimum height=0.8cm,minimum width=8.5cm, below of=c1_11,xshift=4.6cm,yshift=-6.0cm] (on_comb) {Online Combining};
        
        \draw [->] (bcjr12.255) -- (on_comb.173) node [midway] (p1) {$\mathbf{P_1(y)}$} ;
        \draw [->] (bcjrk2.271) -- (on_comb.7) node [midway] (p2k) {$\mathbf{P_M(y)}$};
        
        \node [draw,circle,below of=on_comb,node distance=1.5cm] (p_out) {$\mathbf{\hat{P}(y)}$};
        \draw [->] (on_comb) -- (p_out);
        
    \end{scope}
\end{tikzpicture}
    \caption{Multi-pair BCJR Turbo Decoder}
    \label{fig:multipair}
\end{figure}

During reception,  the decoder pair improves the performance of each other 
by passing the extrinsic information of LLR values. However, if the assumed noise 
distribution for branch transition probabilities is different, the decoding may fail. 
A possible solution to this problem will be to make the decoder pairs  consider 
multiple noise distributions. Towards this end, we propose a multi-pair BCJR decoder 
to consider a wide range of noise distributions.

The proposed multi-pair MAP turbo decoder operates similar to conventional turbo decoder, except
that we have multiple pairs of MAP decoding running in parallel, each for a different noise distribution. 
After multiple rounds of iteration and extrinsic information transfer between
each of the individual decoders, each pair
produces LLR values of the block and the corresponding bit probabilities
$\{\textbf{P}(x_k|\textbf{y})\}_{k=1}^{K}$. 
A schematic representation of the proposed
turbo decoder is provided in Fig. \ref{fig:multipair}.

Let $\mathbf{S}$ denote the systematic bits input to the decoder and  $\mathbf{C_1},
\mathbf{C_2}$ denote the codebits inputs. Traditional turbo decoders have a pair of MAP decoders
denoted by $\mathbf{BCJR_{1}}$ and $\mathbf{BCJR_{2}}$, each acting on $\{\mathbf{S},
\mathbf{C_1}\}$ and $\{\mathbf{S},\mathbf{C_2}\}$ respectively. Let $\mathbf{\Pi}$ and
$\mathbf{\Pi}^{-1}$ denote the interleaver and de-interleaver associated with the code. To
explore multiple noise distributions, we propose $M$-pairs of MAP decoders,
$\left\{ \mathbf{BCJR_{m,1},BCJR_{m,2}} \right\}_{m=1}^{M}$, each designed for a different noise
distribution, $\mathcal{D}_m$, parameterized by $\alpha_m$. Let $\mathbf{P_m}(\mathbf{y})$ denote
the soft information
outputs from each of the decoder pairs. As the channel is unknown, the turbo decoder does not
know which pair of MAP decoder is optimal for the decoding task. Hence, we employ an online
learning framework which ranks each pair of MAP decoders according to its past performance on
the channel and weighs its output $\mathbf{P_m}(\mathbf{y})$ to calculate the final bit
probabilities $\mathbf{\hat{P}}(\mathbf{y})$. 
The computed $\mathbf{\hat{P}}(\mathbf{y})$
can be used to recover the transmitted bits by thresholding at appropriate value.

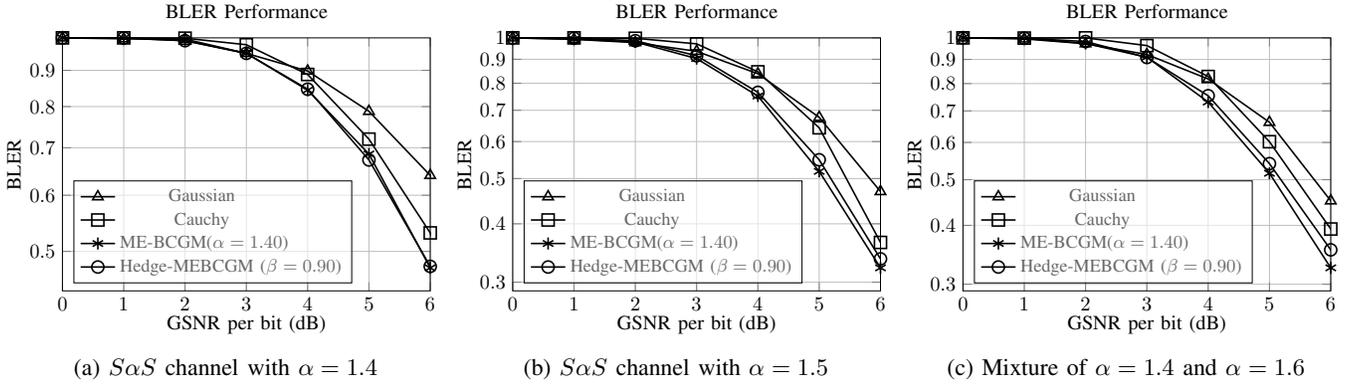
\begin{figure*}[!t]
    \centering
    \begin{subfigure}[t]{.33\linewidth}
        \resizebox{\linewidth}{!}{\begin{tikzpicture}[thick,scale=0.8]
    \begin{semilogyaxis}[
        width=8cm,
        height=6.0cm,
        xmin=0,
        xmax=6,
        ymin=0.00,
        ymax=1.00,
        grid=major,
        xlabel={GSNR per bit (dB)},
        ylabel={BLER},
        xlabel style={at={(0.50,0.05)}},
        ylabel style={at={(0.06,0.50)}},
        ytick={0.00,0.10,...,1.00},
        log ticks with fixed point,
        title={BLER Performance},
        legend pos=south west,
        legend cell align={left},
        legend style={fill opacity=0.6, draw opacity=1.0, text opacity=1.0, minimum width=3cm, font=\small}
        ]
        
        \addplot[black, solid, thick, mark=triangle, mark size={3.0}] 
            table [x=Gaussian_x, y=Gaussian_y, col sep=comma]{./data/data_SaS140_08x10_bler.csv};
        \addlegendentry{Gaussian};

        \addplot[black, solid, thick, mark=square, mark size={3.0}] 
            table [x=Cauchy_x, y=Cauchy_y, col sep=comma]{./data/data_SaS140_08x10_bler.csv};
        \addlegendentry{Cauchy};

        \addplot[black, solid, thick, mark=asterisk, mark size={3.0}] 
            table [x=ME_BCGM_1_40_x, y=ME_BCGM_1_40_y, col sep=comma]{./data/data_SaS140_08x10_bler.csv};
        \addlegendentry{{ME-BCGM($\alpha=1.40$)}};

        \addplot[black, solid, thick, mark=o, mark size={3.0}] 
            table [x=Hegde_MEBCGM_0_90_x, y=Hegde_MEBCGM_0_90_y, col sep=comma]{./data/data_SaS140_08x10_bler.csv};
        \addlegendentry{Hedge-MEBCGM ($\beta=0.90$)};
    \end{semilogyaxis}
\end{tikzpicture}}
        \caption{$S\alpha S$ channel with $\alpha = 1.4$}
        \label{fig:fig_SaS140_08x10_bler}
    \end{subfigure}%
    \begin{subfigure}[t]{.33\linewidth}
        \resizebox{\linewidth}{!}{\begin{tikzpicture}[thick,scale=0.8]
    \begin{semilogyaxis}[
        width=8cm,
        height=6cm,
        xmin=0,
        xmax=6,
        ymin=0.00,
        ymax=1.00,
        grid=major,
        xlabel={GSNR per bit (dB)},
        ylabel={BLER},
        xlabel style={at={(0.50,0.05)}},
        ylabel style={at={(0.06,0.50)}},
        ytick={0.30,0.40,...,1.00},
        log ticks with fixed point,
        title={BLER Performance},
        legend pos=south west,
        legend cell align={left},
        legend style={fill opacity=0.6, draw opacity=1.0, text opacity=1.0, minimum width=3cm, font=\small}
        ]
        
        \addplot[black, solid, thick, mark=triangle, mark size={3.0}] 
            table [x=Gaussian_x, y=Gaussian_y, col sep=comma]{./data/data_SaS150_08x10_bler.csv};
        \addlegendentry{Gaussian};

        \addplot[black, solid, thick, mark=square, mark size={3.0}] 
            table [x=Cauchy_x, y=Cauchy_y, col sep=comma]{./data/data_SaS150_08x10_bler.csv};
        \addlegendentry{Cauchy};

        \addplot[black, solid, thick, mark=asterisk, mark size={3.0}] 
            table [x=ME_BCGM_1_50_x, y=ME_BCGM_1_50_y, col sep=comma]{./data/data_SaS150_08x10_bler.csv};
        \addlegendentry{{ME-BCGM($\alpha=1.40$)}};

        \addplot[black, solid, thick, mark=o, mark size={3.0}] 
            table [x=Hegde_MEBCGM_0_90_x, y=Hegde_MEBCGM_0_90_y, col sep=comma]{./data/data_SaS150_08x10_bler.csv};
        \addlegendentry{Hedge-MEBCGM ($\beta=0.90$)};
    \end{semilogyaxis}
\end{tikzpicture}}
        \caption{$S\alpha S$ channel with $\alpha = 1.5$}
        \label{fig:fig_SaS150_08x10_bler}
    \end{subfigure}%
    \begin{subfigure}[t]{.33\linewidth}
        \resizebox{\linewidth}{!}{\begin{tikzpicture}[thick,scale=0.8]
    \begin{semilogyaxis}[
        width=8cm,
        height=6cm,
        xmin=0,
        xmax=6,
        ymin=0.00,
        ymax=1.00,
        grid=major,
        xlabel={GSNR per bit (dB)},
        ylabel={BLER},
        xlabel style={at={(0.50,0.05)}},
        ylabel style={at={(0.06,0.50)}},
        ytick={0.30,0.40,...,1.00},
        log ticks with fixed point,
        title={BLER Performance},
        legend pos=south west,
        legend cell align={left},
        legend style={fill opacity=0.6, draw opacity=1.0, text opacity=1.0, minimum width=3cm, font=\small}
        ]
        
        \addplot[black, solid, thick, mark=triangle, mark size={3.0}] 
            table [x=Gaussian_x, y=Gaussian_y, col sep=comma]{./data/data_SaSMix140160050_bler.csv};
        \addlegendentry{Gaussian};

        \addplot[black, solid, thick, mark=square, mark size={3.0}] 
            table [x=Cauchy_x, y=Cauchy_y, col sep=comma]{./data/data_SaSMix140160050_bler.csv};
        \addlegendentry{Cauchy};

        \addplot[black, solid, thick, mark=asterisk, mark size={3.0}] 
            table [x=MEBCGM_for_Mixture_x, y=MEBCGM_for_Mixture_y, col sep=comma]{./data/data_SaSMix140160050_bler.csv};
        \addlegendentry{{ME-BCGM($\alpha=1.40$)}};

        \addplot[black, solid, thick, mark=o, mark size={3.0}] 
            table [x=Hegde_MEBCGM_0_90_x, y=Hegde_MEBCGM_0_90_y, col sep=comma]{./data/data_SaSMix140160050_bler.csv};
        \addlegendentry{Hedge-MEBCGM ($\beta=0.90$)};
    \end{semilogyaxis}
\end{tikzpicture}}
        \caption{Mixture of $\alpha = 1.4$ and $\alpha = 1.6$}
        \label{fig:fig_SaSMix140160050_08x10_bler}
    \end{subfigure}%
    \caption{BLER Comparison of proposed method}
    \label{fig:bler_curve}
\end{figure*}

Considering each pair of MAP decoders as an \textit{expert}, the task of combining the individual
results can be viewed as the problem of \textit{prediction with multiple experts} from game
theory. Ideally, more importance should be given to those decoder pairs which are able to decode the past
blocks successfully. However, due to the presence of noise in the received signal, the decoding
performance on the individual pairs will also be noisy. Motivated by these constraints, we 
use Hedge algorithm \cite{freund1997decision} for the online combining of the different MAP 
decoder pair predictions. 

Hedge is an online learning algorithm for combining
decisions from multiple experts. It considers decisions from $M$
experts at each time step and combines these observations based on the past performance of
the experts. A learning parameter $\beta \in (0,1]$ is used by the algorithm to update the
importance it gives to each of the experts after observing the loss for each expert. The 
learning parameter regulates how the instantaneous observations affect the importance of
every expert at each timestep. A high value of $\beta$  will make the algorithm strongly
resilient to noises in the observation of loss at the cost of slowing down the learning procedure.
On the other hand a low value of $\beta$ will cause the algorithm to react quickly to observation
noises but at the cost of ignoring the past performances. The optimal value
of $\beta$ to use is problem dependent. In \cite{freund1997decision}, the optimal value
of $\beta$ to use under the worst case guarantees is provided.

\begin{algorithm}
    \begin{algorithmic}[1]
        \caption{Multi-pair MAP Decoding for Turbo Codes}    \label{alg:hedge_turbo}
        \State \textbf{Parameters: } $ \beta \in (0,1]$, $M \in I^+$ 
        \State \textbf{Initialization: } Set $w_m(1) = W > 0 \:\forall\: m$
        \For{ each new block of received data }
            \State Get $\mathbf{P_m}(\mathbf{y}) \: \forall \:m$, from each MAP decoders
            \For{ $m = 1,\ldots,M$ }
                \State $\zeta_m \gets \frac{w_m}{\sum \limits_{j=1}^{M}w_j}$  \label{eqn:normWeighCalc}
            \EndFor
            \State Calculate $\mathbf{\hat{P}}(\mathbf{y})$ as $\mathbf{\hat{P}}(\mathbf{y}) = \sum 
                        \limits_{m=1}^{M} \zeta_m \mathbf{P_m}(\mathbf{y})$
            \State Get block decoding loss $l_m \: \forall \: m$ 
            \State Update weights as $w_{m} \gets w_m \cdot \beta^{l_m}$  \label{eqn:weightUpdate}
        \EndFor
    \end{algorithmic}
\end{algorithm}

The proposed online learning algorithm for decoding is given in Alg. \ref{alg:hedge_turbo}.
The algorithm works by assigning a weight $W$ to each
decoder pair and then updating the weights based on the observed block decoding performance.
These weights are normalized (Line 6 of Alg. \ref{alg:hedge_turbo}) and bit
probabilities $\{\textbf{P}_m(\textbf{y})\}_{m=1}^{M}$ produced by each of the decoder pairs are
combined using these normalized weights $\{\zeta_m\}_{m=1}^{M}$ (Line 8)
to compute the final bit probabilities  $\mathbf{\hat{P}}(\mathbf{y})$.
For each MAP decoder pair, the computed bit probabilities
$\{\textbf{P}_m(\textbf{y})\}_{m=1}^{M}$ can be used by any block code error detection mechanism
(like Cyclic Redundancy Check (CRC)) to check whether the decoder pair is able to perform the
block decoding successfully. The loss value $l_m \in [0,1]$ is calculated for each
decoder pair $m = 1, \ldots, M$ where $l_m$ is the fraction of bit errors in the block. 
Finally, the weights of individual decoders are updated based on an exponential rule (Line 10). 
    \section{Results}   \label{sec:results}
To validate the usefulness of the proposed approach, we consider a binary antipodal signaling
system with turbo coding in an impulsive noisy channel modeled using $S \alpha S $ with $\alpha$
unknown to the receiver. Turbo code blocks are generated using two Recursive  Systematic
Convolutional (RSC) encoders linked by a random interleaver. The encoders are defined by 
polynomials $1+D+D^2$ and $1+D^2$, with $1+D+D^2$ as feedback polynomial and a constraint length of
$4$. Puncturing is used to obtain higher rates. We provide results for a coding rate of $4/5$.
Input sequence are split into blocks $\textbf{d} = \{d_1,d_2,\ldots,d_{128}\}$ of length $K = 128$
and are encoded to bipolar code sequences of block length $172$ with $12$ tail bits for parking the
encoder.

We used multi-pair MAP turbo decoder with  $M = 6$ pairs of BCJR decoders. The distributions
$\mathcal{D}_k$ are selected to be $\mathcal{D}_1 = Cauchy(\gamma)$, $\{\mathcal{D}_k\}_{k=2}^{5} =
\{S \alpha S(0.2 \times (k-2) +1.2,\gamma)_{k=2}^{5}$\} and $\mathcal{D}_6 =
\mathcal{N}(0,2\gamma)$. Even though we chose an evenly spaced parameter range of $\alpha$, any set of
distributions can also be used, if such prior information is available. Each pair of decoders  
ran for $8$ iterations for decoding each block. 
For learning algorithm, we used $W = 1.0$ and $\beta = 0.90$.
Experiments are
conducted for channel conditions which can be optimally decoded 
\begin{enumerate}
    \item by at least one of the decoder pairs in the pool,
    \item with none of the decoders in the pool,
    \item with a mixture of two decoders of $S\alpha S$ noises.
\end{enumerate}

\subsection{Performance Evaluation}
The Block Error Rate (BLER) performance obtained is provided in Fig. \ref{fig:bler_curve}. We compare the performance
of the proposed approach with the performance of Gaussian and Cauchy detectors (for which closed-form
expressions of density are available) as well as the optimal decoding based on ME-BCGM\cite{xu2012minimum},
which assumes the knowledge of $\alpha$ known apriori (using (\ref{eqn:eps_quad_fit})
in (\ref{eqn:bcgm_model})). For the proposed method, the reported values are obtained over
a combined online training-testing period of $10000$ transmission blocks. For other methods, the reported
values are obtained using Monte Carlo simulations until $50$ block errors are obtained or at least $10000$
blocks are transmitted, whichever occurs last.

Fig. \ref{fig:fig_SaS140_08x10_bler} shows the scenarios where the optimal decoder for the observed
channel is present in the pool of experts. We used $\alpha = 1.4$ similar to \cite{dimanche2014detection,mestrah2018blind}. 
We can see that the proposed method, without any knowledge
about the channel impulsiveness parameter $\alpha$, is able to match the BLER performance of the optimal
decoder which has the knowledge of parameter $\alpha$. In Fig. \ref{fig:fig_SaS150_08x10_bler}, we 
provide the results for a channel when the optimal decoder is not available in the pool of experts. We
used a $S\alpha S$ channel with $\alpha = 1.5$, but the multi-pair decoder did not include the 
corresponding optimal decoder. It can be observed that the proposed method is able to closely follow the
performance of the optimal decoder which has the perfect knowledge of impulsiveness factor $\alpha$.
In Fig. \ref{fig:fig_SaSMix140160050_08x10_bler}, the performance comparison under a mixed noise model
channel is given. The channel is modeled using two noise components of equal strength: $\alpha = 1.4$
and $\alpha = 1.6$. When compared with the optimal decoder (labeled as 'ME-BCGM for Mixture'), the 
proposed method provides a competitive BLER performance. 

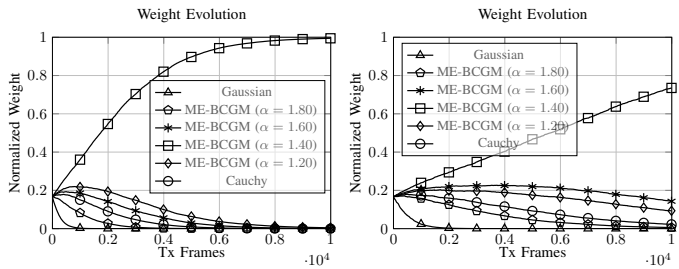
\begin{figure}[!t]
    \centering
    \begin{subfigure}{.25\textwidth}
        \resizebox{\linewidth}{!}{\begin{tikzpicture}[thick]
    \begin{axis}[
        width=8cm,
        height=6cm,
        xmin=0,
        xmax=10000,
        ymin=0.00,
        ymax=1.00,
        grid=major,
        xlabel={Tx Frames},
        ylabel={Normalized Weight},
        xlabel style={at={(0.50,0.05)}},
        ylabel style={at={(0.06,0.50)}},
        ytick={0.00,0.20,...,1.00},
        log ticks with fixed point,
        title={Weight Evolution},
        legend cell align={left},
        legend style={fill opacity=0.6, draw opacity=1.0, text opacity=1.0, minimum width=3cm, font=\small, at={(0.98,0.48)}, anchor=east}
        ]
        
        \addplot[black, solid, thick, mark=triangle, mark size={3.0}, mark repeat=100, mark phase=100] 
            table [x=Gaussian_x, y=Gaussian_y, col sep=comma]{./data/data_SaS140_wevo.csv};
        \addlegendentry{Gaussian};

        \addplot[black, solid, thick, mark=pentagon, mark size={3.0}, mark repeat=100, mark phase=100] 
            table [x=ME_BCGM_1_80_x, y=ME_BCGM_1_80_y, col sep=comma]{./data/data_SaS140_wevo.csv};
        \addlegendentry{ME-BCGM ($\alpha=1.80$)};

        \addplot[black, solid, thick, mark=asterisk, mark size={3.0}, mark repeat=100, mark phase=100] 
            table [x=ME_BCGM_1_60_x, y=ME_BCGM_1_60_y, col sep=comma]{./data/data_SaS140_wevo.csv};
        \addlegendentry{ME-BCGM ($\alpha=1.60$)};

        \addplot[black, solid, thick, mark=square, mark size={3.0}, mark repeat=100, mark phase=100] 
            table [x=ME_BCGM_1_40_x, y=ME_BCGM_1_40_y, col sep=comma]{./data/data_SaS140_wevo.csv};
        \addlegendentry{ME-BCGM ($\alpha=1.40$)};

        \addplot[black, solid, thick, mark=diamond, mark size={3.0}, mark repeat=100, mark phase=100] 
            table [x=ME_BCGM_1_20_x, y=ME_BCGM_1_20_y, col sep=comma]{./data/data_SaS140_wevo.csv};
        \addlegendentry{ME-BCGM ($\alpha=1.20$)};

        \addplot[black, solid, thick, mark=o, mark size={3.0}, mark repeat=100, mark phase=100] 
            table [x=Cauchy_x, y=Cauchy_y, col sep=comma]{./data/data_SaS140_wevo.csv};
        \addlegendentry{Cauchy};
    \end{axis}
\end{tikzpicture}}
        \caption{$S\alpha S$ channel with $\alpha = 1.4$}
        \label{fig:fig_SaS140_08x10_wevo}
    \end{subfigure}%
    \begin{subfigure}{.25\textwidth}
        \resizebox{\linewidth}{!}{\begin{tikzpicture}[thick]
    \begin{axis}[
        width=8cm,
        height=6cm,
        xmin=0,
        xmax=10000,
        ymin=0.00,
        ymax=1.00,
        grid=major,
        xlabel={Tx Frames},
        ylabel={Normalized Weight},
        xlabel style={at={(0.50,0.05)}},
        ylabel style={at={(0.06,0.50)}},
        ytick={0.00,0.20,...,1.00},
        log ticks with fixed point,
        title={Weight Evolution},
        legend cell align={left},
        legend pos=north west,
        legend style={fill opacity=0.6, draw opacity=1.0, text opacity=1.0, minimum width=3cm, font=\small}
        ]
        
        \addplot[black, solid, thick, mark=triangle, mark size={3.0}, mark repeat=100, mark phase=100] 
            table [x=Gaussian_x, y=Gaussian_y, col sep=comma]{./data/data_SaSMix140160_wevo.csv};
        \addlegendentry{Gaussian};

        \addplot[black, solid, thick, mark=pentagon, mark size={3.0}, mark repeat=100, mark phase=100] 
            table [x=ME_BCGM_1_80_x, y=ME_BCGM_1_80_y, col sep=comma]{./data/data_SaSMix140160_wevo.csv};
        \addlegendentry{ME-BCGM ($\alpha=1.80$)};

        \addplot[black, solid, thick, mark=asterisk, mark size={3.0}, mark repeat=100, mark phase=100] 
            table [x=ME_BCGM_1_60_x, y=ME_BCGM_1_60_y, col sep=comma]{./data/data_SaSMix140160_wevo.csv};
        \addlegendentry{ME-BCGM ($\alpha=1.60$)};

        \addplot[black, solid, thick, mark=square, mark size={3.0}, mark repeat=100, mark phase=100] 
            table [x=ME_BCGM_1_40_x, y=ME_BCGM_1_40_y, col sep=comma]{./data/data_SaSMix140160_wevo.csv};
        \addlegendentry{ME-BCGM ($\alpha=1.40$)};

        \addplot[black, solid, thick, mark=diamond, mark size={3.0}, mark repeat=100, mark phase=100] 
            table [x=ME_BCGM_1_20_x, y=ME_BCGM_1_20_y, col sep=comma]{./data/data_SaSMix140160_wevo.csv};
        \addlegendentry{ME-BCGM ($\alpha=1.20$)};

        \addplot[black, solid, thick, mark=o, mark size={3.0}, mark repeat=100, mark phase=100] 
            table [x=Cauchy_x, y=Cauchy_y, col sep=comma]{./data/data_SaSMix140160_wevo.csv};
        \addlegendentry{Cauchy};
    \end{axis}
\end{tikzpicture}}
        \caption{Mixture of $\alpha = 1.4$ and $1.6$}
        \label{fig:fig_SaSMix140160_wevo}
    \end{subfigure}%
    \caption{Weight Evolution of experts at different channels}
    \label{fig:wevo_curve}
\end{figure}

The weight evolution of each expert during the simulation period for the different scenarios is given in
Fig. \ref{fig:wevo_curve}. 
From the weight evolution of the experts provided in Fig. \ref{fig:wevo_curve},
we can see that the Gaussian decoder, which corresponds to an $\alpha$ value of $2$, is getting suppressed
very fast in all the cases. When the channel is $\alpha = 1.4$ (shown in Fig.
\ref{fig:fig_SaS140_08x10_wevo}), the ME-BCGM decoder which uses $\alpha =  1.4$ is getting the 
highest weight as
expected. 
In the scenario with mixture of two noises (\ref{fig:fig_SaSMix140160_wevo}), the weight
evolution is slow, but eventually the decoder with $\alpha = 1.4$ dominates as time 
progresses, followed by the decoder with $\alpha = 1.6$.

\subsection{Reducing Computational Complexity}
The price we pay for improving the decoding performance
in the proposed scheme is the increase in computational complexity. Because multiple decoder pairs
are required to compute LLRs under different candidate distributions, the computational complexity
of the proposed solution increases linearly with the number of experts considered. 

One solution to reduce the computational complexity of the proposed method is to shutdown
some of the decoder pairs once enough confidence is gained by the learning algorithm about
their performance. The normalized weight $\zeta_m$ for each of the decoder pair can be seen
as a quantitative metric about its performance. This information can be
used to selectively shutdown decoder pairs. To demonstrate this, we repeated the above
experiment, but with the early stopping of weight updates at timestep $\tau$ and picking 
the best decoder-pair (with highest normalized weight) at that instant for decoding 
the rest of the messages. 
A comparison BLER
performance for different values of $\tau$ is provided in
Table \ref{tab:early_stop} for a transmission of $100000$ blocks. 

\begin{table}[h]
    \centering
    \begin{tabular}{|r|r|r|r|r|}
        \hline
        \multicolumn{1}{|c|}{\multirow{2}{*}{$\tau$}} 
            & \multicolumn{2}{c|}{GNSR = $10$dB} 
            & \multicolumn{2}{c|}{GSNR = $12$dB} \\ \cline{2-5} 
        \multicolumn{1}{|c|}{}                        
            & \multicolumn{1}{c|}{$\alpha = 1.40$} 
            & \multicolumn{1}{c|}{$\alpha = 1.50$} 
            & \multicolumn{1}{c|}{$\alpha = 1.40$} 
            & \multicolumn{1}{c|}{$\alpha = 1.50$} \\ \hline
        500  & $0.0547$ & $0.0271$ & $0.0169$ & $0.0074$ \\
        1000 & $0.0514$ & $0.0261$ & $0.0151$ & $0.0071$ \\
        1500 & $0.0523$ & $0.0241$ & $0.0156$ & $0.0068$ \\
        2000 & $0.0526$ & $0.0249$ & $0.0132$ & $0.0070$ \\
        2500 & $0.0498$ & $0.0250$ & $0.0131$ & $0.0068$ \\
        \hline
    \end{tabular}
    \caption{Effect of early stopping in BLER.} \label{tab:early_stop}
\end{table}

We can observe that an early stopping of weight update and choosing the decoder-pair
with high $\zeta_m $ is not affecting the BLER by a huge margin. As the $\tau$ increases,
we can see a trend of decreasing BLER. This is because, as $\tau$ increases, the normalized
weight of best decoder-pair also increases and this help in confidently selecting
the best decoder-pair for the channel conditions. This reduces the computational
complexity of the proposed method to that of the traditional turbo decoding scheme
after $\tau$ timesteps.

\subsection{Effect of learning parameter $\bm{\beta}$}
The learning parameter 
$\beta$ decides how fast or slow the hedge algorithm responds to errors of the individual experts. 
For a low value of $\beta$, the final decision will be susceptible to noises in the individual expert
decisions. A high value of $\beta$ will make the algorithm delay the boosting of the best expert.
Even though an expression for the value of $\beta$ for provable
loss guarantees under worst-case scenarios is derived in \cite{freund1997decision}, in practice 
it is found that the
value of the $\beta$ that gives the best performance depends on the particular scenario under test. 
In our experiments,
we found the setting the value of $\beta$ in the range of $[0.85,0.99]$ gives almost the same performance
in terms of BLER. This suggests that the proposed method can be used without extensive parameter tuning with
almost no loss in performance.

    \section{Concluding Remarks}
In this paper, we introduced a novel method for performing error correction coding 
in the absence of knowledge about the parameters of the noise distribution
through online learning. By combining decisions 
from multiple decoders, we showed that the performance close to the optimal receiver can
be obtained without any prior knowledge about the noise parameters.
In this work, we chose the pool of decoder-pairs with $\alpha$
parameter linearly spaced between $1$ and $2$. An interesting future direction can be to explore
the number of decoder-pairs to be used in the pool and the values of $\alpha$s to choose. 
    
    \bibliographystyle{IEEEtran}
    \bibliography{library.bib}

\end{document}